\documentclass{article}
\usepackage{spconf,amsmath,graphicx}
\usepackage[dvipsnames]{xcolor}

\usepackage{amssymb}
\usepackage{amsbsy}
\usepackage{subcaption}
\newcommand\blfootnote[1]{%
  \begingroup
  \renewcommand\thefootnote{}\footnote{#1}%
  \addtocounter{footnote}{-1}%
  \endgroup
}

\ninept
\title{A COMPARISON OF POOLING METHODS ON LSTM MODELS\\FOR RARE ACOUSTIC EVENT CLASSIFICATION}
\name{Chieh-Chi Kao$^1$, Ming Sun$^1$, Weiran Wang$^{2*}$, Chao Wang$^1$}
\address{$^1$Amazon.com Inc. $\qquad$ $^2$Salesforce Research\\
\texttt{\small{\{chiehchi,mingsun\}@amazon.com $\qquad$ weiran.wang@salesforce.com} $\qquad$ wngcha@amazon.com}}
\begin{document}
%
\maketitle
\begin{abstract}
Acoustic event classification (AEC) and acoustic event detection (AED) refer to the task of detecting whether specific target events occur in audios.
As long short-term memory (LSTM) leads to state-of-the-art results in various speech related tasks, it is employed as a popular solution for AEC as well.
This paper focuses on investigating the dynamics of LSTM model on AEC tasks. 
It includes a detailed analysis on LSTM memory retaining, and a benchmarking of nine different pooling methods on LSTM models using 1.7M generated mixture clips of multiple events with different signal-to-noise ratios. 
This paper focuses on understanding: 1)~utterance-level classification accuracy; 2) sensitivity to event position within an utterance.
The analysis is done on the dataset for the detection of rare sound events from DCASE 2017 Challenge.
We find max pooling on the prediction level to perform the best among the nine pooling approaches in terms of classification accuracy and insensitivity to event position within an utterance.
To authors' best knowledge, this is the first kind of such work focused on LSTM dynamics for AEC tasks.

\end{abstract}
\section{Introduction}
\label{s:intro}
\blfootnote{$^*$This work was done while the author was at Amazon.}
Acoustic event classification (AEC) and acoustic event detection (AED) refer to the task of detecting whether specific target events occur in audios. 
It is of interest in many real-world scenarios, e.g. traffic monitoring \cite{chen1997automatic}, surveillance \cite{cristani2007audio, valenzise2007scream}, etc. 
In recent years, with deep learning based solutions burgeoning in various areas including speech and language processing \cite{hinton2012deep, Sundermeyer_lstmneural}, and computer vision \cite{Byeon_2015_CVPR}, there has been intensive research on improving AED performance via neural networks. 
For instance, several new neural network architectures \cite{kao2018r,wang2018_IS,wang2019comparison} were proposed for AED. 

As long short-term memory (LSTM)~\cite{hochreiter1997long} leads to state-of-the-art results in various speech related tasks, e.g. automatic speech recognition~\cite{hinton2012deep}, keyword spotting~\cite{Arik2017}, speaker identification~\cite{ren2016look}, whisper detection~\cite{Raeesy2018SLT}, it is employed as a popular solution for AEC as well~\cite{LSTM_SED,wang2016MED,Parascandolo,Guo2017,Tang2018,Shi2018,Shi2019}, typically combined with convolutional neural networks (CNNs) \cite{Lim2017}. 
To run applications mentioned above on mobile devices or smart speakers, a model with small memory footprint is required.
LSTM models have much less number of parameters than CNN models while reasonable performance maintained.
Besides, LSTM models can operate in a streaming mode with a ring buffer, which has the benefit of low latency.
We are interested in understanding the memory dynamics for LSTM based AEC models, i.e. for how long LSTM retains memory of happening acoustic events, as well as how to improve memory retaining for specific pooling methods. Recent research investigates memory dynamics and control in recurrent neural networks (RNNs) including LSTM \cite{haviv2019understanding}. There are comparisons on different pooling functions for AEC/AED~\cite{wang2019comparison}, and spatio-temporal attention pooling proposed for audio scene classification~\cite{phan2019pooling}.

\begin{figure}[t!]
    \centering
    \begin{subfigure}[b]{0.2\textwidth}
        \includegraphics[width=\textwidth]{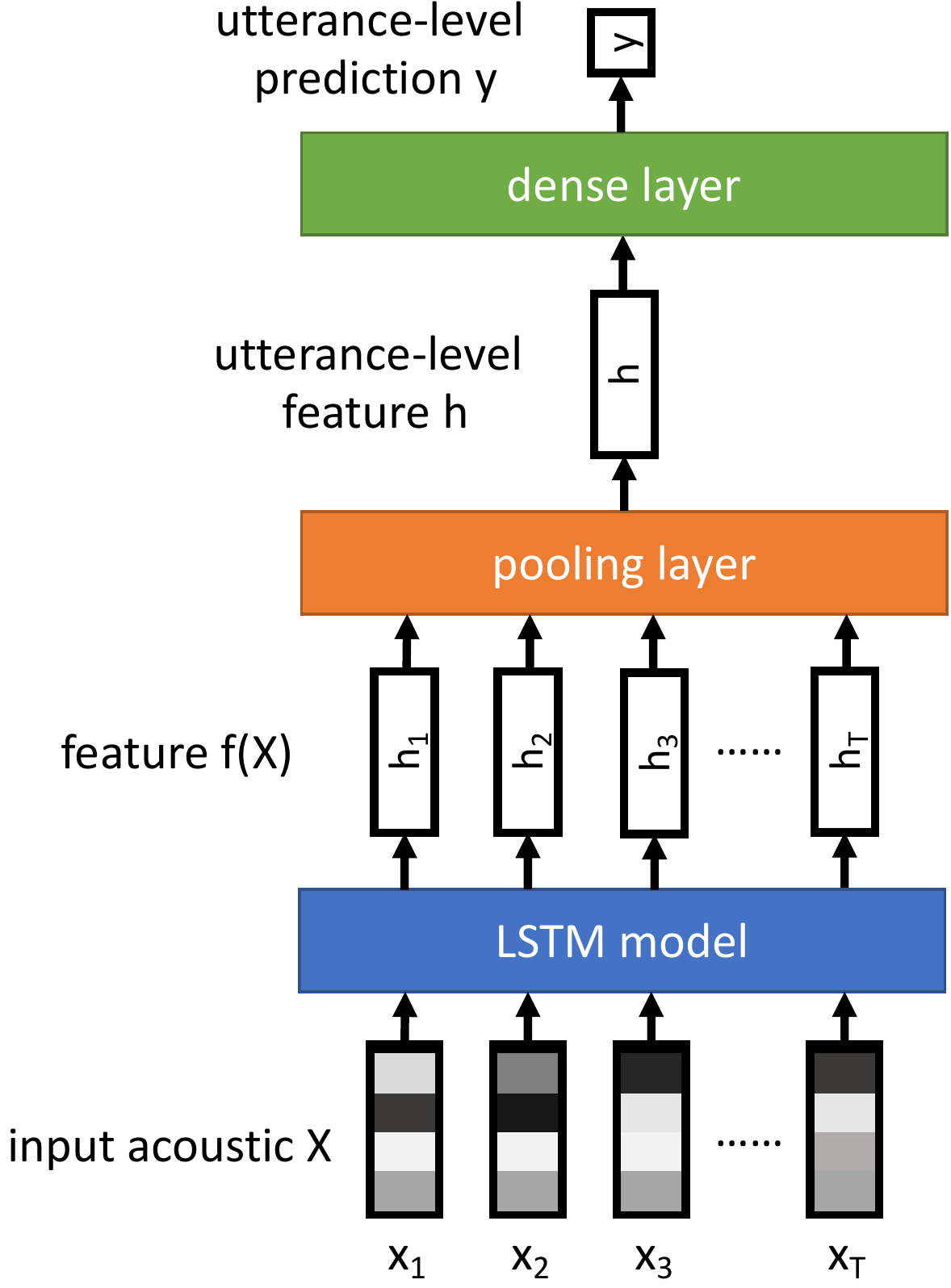}
        \caption{Pooling on feature}
        \label{fig:pool_on_feat}
    \end{subfigure}
    ~ 
    \begin{subfigure}[b]{0.2\textwidth}
        \includegraphics[width=\textwidth]{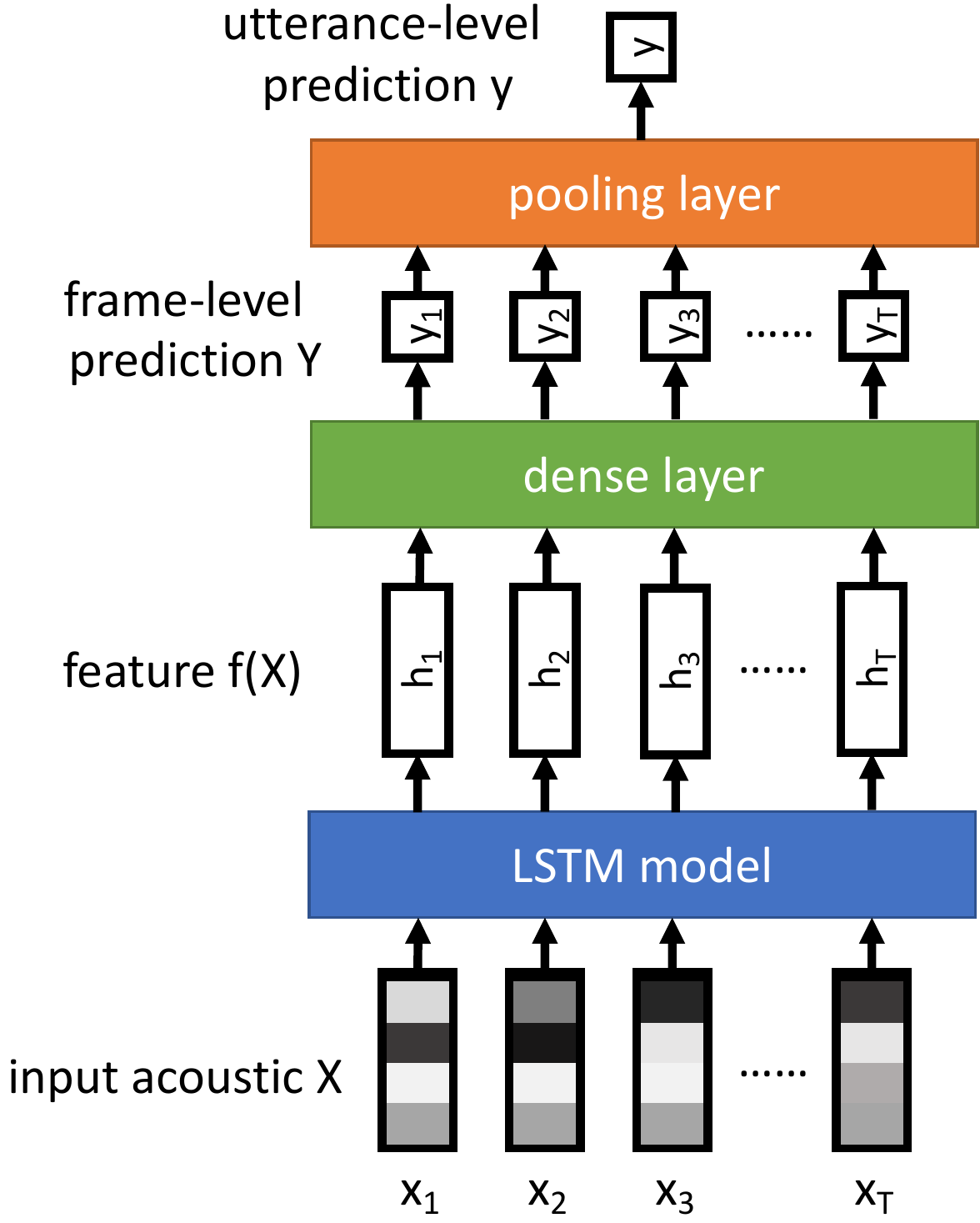}
        \caption{Pooling on prediction}
        \label{fig:pool_on_pred}
    \end{subfigure}
    \caption{Two categories of pooling methods to generate utterance-level prediction for LSTM models.}
    \label{fig:pooling_types}
    \vspace{-0.2in}
\end{figure}

Wang et al.~\cite{wang2019comparison} did a thorough analysis theoretically and experimentally of five pooling functions on prediction.
The analysis was done for multiple instance learning framework on AED with weak labeling, whose goal is to detect and localize events at the same time.
Their experiments were done on DCASE 2017 task~4~\cite{DCASE2017challenge}: weakly supervised AED for smart cars.
This dataset employs a subset of AudioSet~\cite{AudioSet}, which includes 10-second clips containing 17 sound events from two categories: ``Warning'' and ``Vehicle''.
Although audio tagging, which is similar to our utterance-level classification task, is covered in the experiments in~\cite{wang2019comparison}, there are two fundamental differences between this work and theirs.
1) \textbf{Different characteristics of events}: 
Our work focuses on rare events, and we conducted experiments on DCASE 2017 task~2~\cite{DCASE2017challenge}: detection of rare sound events. 
The averaged length of target events in the test set is shorter than 2 seconds~\cite{wang2017_dcaseT2}, and these events only occupy a very small portion of the whole utterance (30 seconds).
On the contrary, for events in DCASE 2017 task~4 (e.g. ``Ambulance (siren)'', ``Car passing by'', ``Train'', etc.), the sound may cover the whole 10-second clip. 
2) \textbf{Different focuses of analysis}:
In~\cite{wang2019comparison}, the analysis focuses on the effect of pooling functions on both event localization and classification for weak labeling.
Due to the requirement of event localization, only pooling functions on prediction were discussed in~\cite{wang2019comparison}.
Our work analyzes the effect of pooling functions on LSTM based AEC models focusing on utterance-level classification. 
Experiments are designed for understanding the memory dynamics for LSTM models, and looking for solutions to mitigate the sensitivity to event positions.
Moreover, we further discuss four pooling functions on the feature side, which are not covered in~\cite{wang2019comparison}.

In this paper, we investigate the dynamics of LSTM memory on AEC tasks, including an analysis on LSTM memory retaining, and a benchmarking for impacts of different pooling approaches on LSTM memory dynamics and AEC accuracies, using 1.7M synthesized clips (across 3 event types and 3 SNRs). 
We also show that bi-directional LSTM can mitigate the sensitivity to event positions for certain pooling methods.

\begin{table}
\begin{center}
\caption[Table caption text]{Pooling methods on feature.}
\label{t:poolingFeat}
\vspace{-0.1in}
\begin{tabular}{ | l | l |}
\hline
Pooling method & Function   \\ \hline
Last frame & $\boldsymbol{h}=\boldsymbol{h}_T$  \\ \hline
Attention & $\boldsymbol{h}=\sum_{t}{a_t}\boldsymbol{h}_t$   \\ \hline
Max pooling & $\boldsymbol{h}[n]=\max_{t}\boldsymbol{h}_t[n]$ \\ \hline
Average pooling & $\boldsymbol{h}=\frac{1}{T}\sum_{t}\boldsymbol{h}_t$  \\ \hline
\end{tabular}
\vspace{-0.2in}
\end{center}
\end{table}

\begin{table}
\begin{center}
\caption[Table caption text]{Pooling methods on prediction.}
\label{t:poolingPred}
\vspace{-0.1in}
\begin{tabular}{ | l | l |}
\hline
Pooling method & Function   \\ \hline
Max pooling & $y=\max_{t}y_t$ \\ \hline
Average pooling & $y=\frac{1}{T}\sum_{t}y_t$  \\ \hline
Linear softmax & $y=\frac{\sum_{t}{y_t}^2}{\sum_{t}{y_t}}$  \\ \hline
Exponential softmax & $y=\frac{\sum_{t}{y_t}\exp{(y_t)}}{\sum_{t}{\exp{(y_t)}}}$   \\ \hline
Attention & $y=\frac{\sum_{t}{y_t}{w_t}}{\sum_{t}{w_t}}$ \\ \hline
\end{tabular}
\vspace{-0.2in}
\end{center}
\end{table}
\vspace{-0.2in}

\section{Pooling functions for LSTM models}
\label{s:poolingFunctions}

Denote an input utterance by $\boldsymbol{X}$=[$\boldsymbol{x}_1$,...,$\boldsymbol{x}_T$] where $\boldsymbol{x}_i \in \mathbb{R}^d$ contains the audio features for the $i$-th frame. 
Since our task is to detect whether the specific target event occurs in the utterance, a binary label $\hat{y}$ which indicates if an event occurs ($\hat{y}$ = 1) or not ($\hat{y}$ = 0) is given for each utterance. 
Our goal is to make accurate predictions at utterance-level for rare acoustic event classification.

Our model uses an LSTM model $f$, which may contain one or multiple LSTM layers, to extract nonlinear features from $\boldsymbol{X}$. 
This gives a representation feature containing temporal information as: $f(\boldsymbol{X}) = [\boldsymbol{h}_1,...,\boldsymbol{h}_T] \in \mathbb{R}^{N\times{T}}$.
Given this feature $f(\boldsymbol{X})$, different pooling methods can be applied to generate the utterance-level prediction $y$.
We can categorize pooling methods into two types: \textit{pooling on feature} and \textit{pooling on prediction}.

\subsection{Pooling on feature}
\label{s:poolingFeat}
Given the feature $f(\boldsymbol{X})$ generated by an LSTM model, we can aggregate the frame-level features to generate the utterance-level feature $\boldsymbol{h}$.
The utterance-level feature is then fed into a dense layer with sigmoid activation to generate the utterance-level prediction $y$.
We list the definition of four feature-level pooling functions in Table~\ref{t:poolingFeat}.

Using the feature of the last frame in the utterance ($\boldsymbol{h}_T$) is the simplest way to generate the utterance-level feature.
Since LSTM models have the ability to capture contextual information, we can directly use the feature of the last frame to represent the whole utterance. 
However, this na\"ive method may suffer from long-term memory loss.
Li et al.~\cite{Li_2018_CVPR} showed that LSTM models could only memorize less than 1,000 steps.
Our experiments shown in Sec.~\ref{s:dynamics} also have demonstrated that this method has the forgetting issue on AEC.

The attention pooling function uses attention weights ($a_t$) to combine frame-level features to form the utterance-level feature.
We use the same architecture as proposed in~\cite{wang2018_IS} to generate attention weights.
The weights for each frame ($a_t$) are learned with a dense layer, whose weights are shared with the final dense layer generating the utterance-level prediction.
This design encourages the attention to be peaked at frames containing target events, and be suppressed at the non-event frames.
In this work, there is no frame-level supervision for attention, which is slightly different from~\cite{wang2018_IS}.

The max pooling function takes the largest value at each feature channel across all time frames.
Intuitively, it should work well for rare event detection since the length of target event is short compared to the utterance length.
Once the model detects the event at certain frames, these strong responses will be kept through max pooling without suffering from the forgetting issue.
Our experiments also have shown that this is one of the best performing pooling methods.

The average pooling function simply assigns the same weight ($\frac{1}{T}$) to all frames. 
This setup is not suitable for rare AEC since the number of frames with events is small, which makes the utterance-level feature difficult to represent the events.


\begin{figure}[t!]
    \centering
    \begin{subfigure}[b]{0.38\textwidth}
        \includegraphics[width=\textwidth]{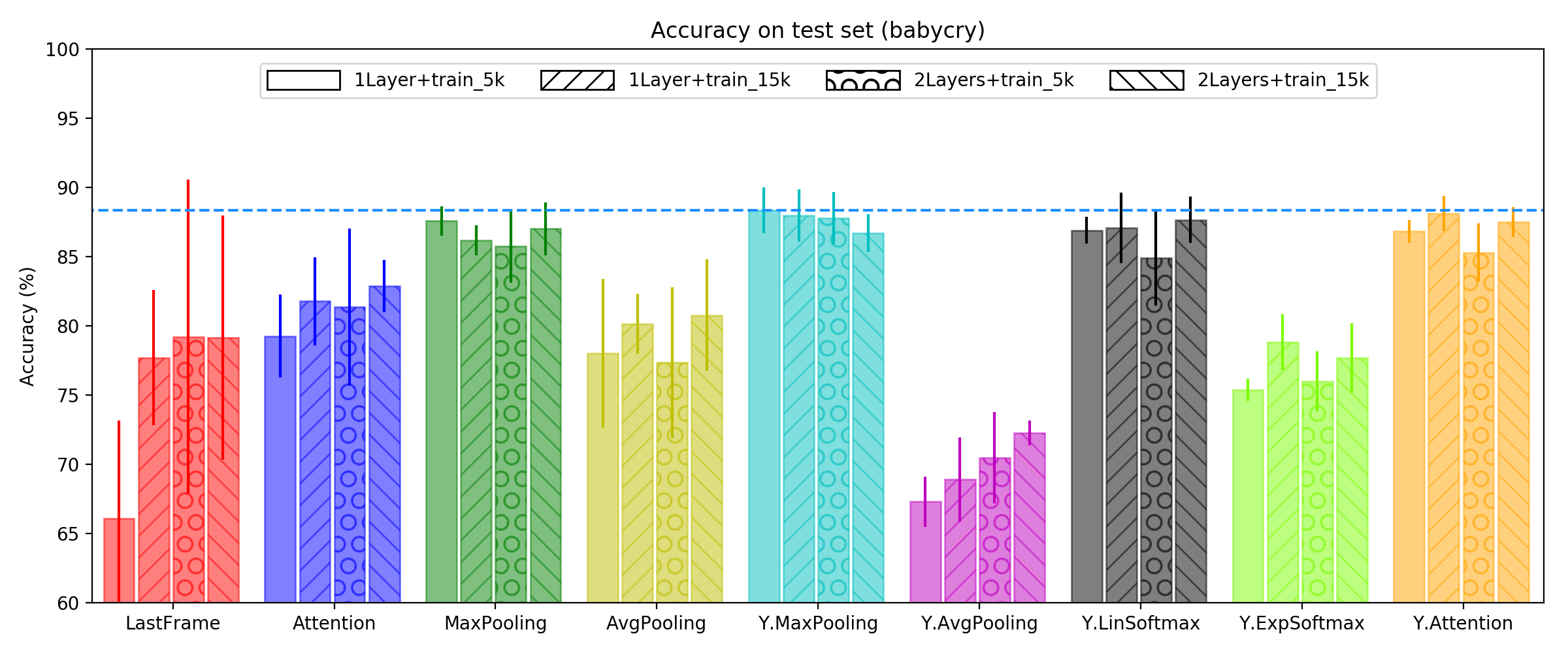}
        \caption{Babycry}
        \label{fig:test_acc_babycry}
    \end{subfigure}
    ~ 
    \begin{subfigure}[b]{0.38\textwidth}
        \includegraphics[width=\textwidth]{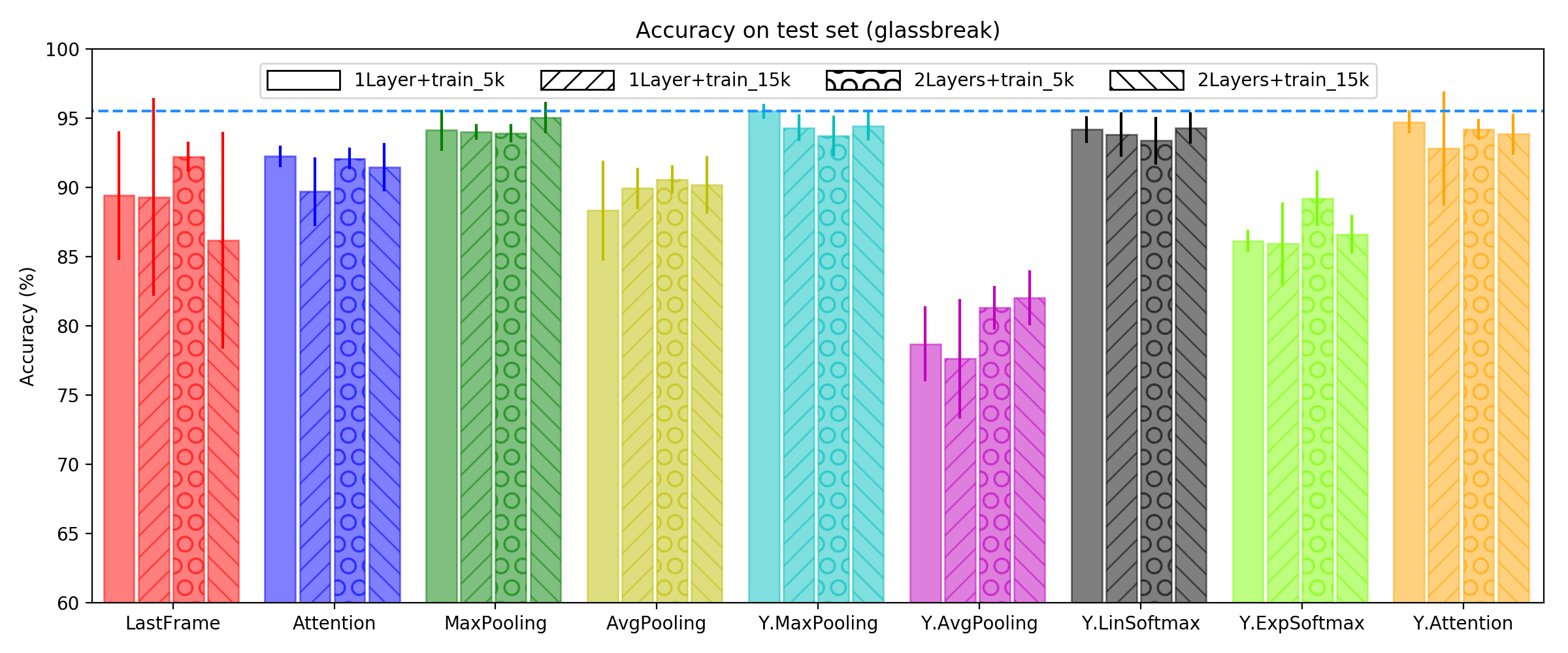}
        \caption{Glassbreak}
        \label{fig:test_acc_glassbreak}
    \end{subfigure}
    ~ 
    \begin{subfigure}[b]{0.38\textwidth}
        \includegraphics[width=\textwidth]{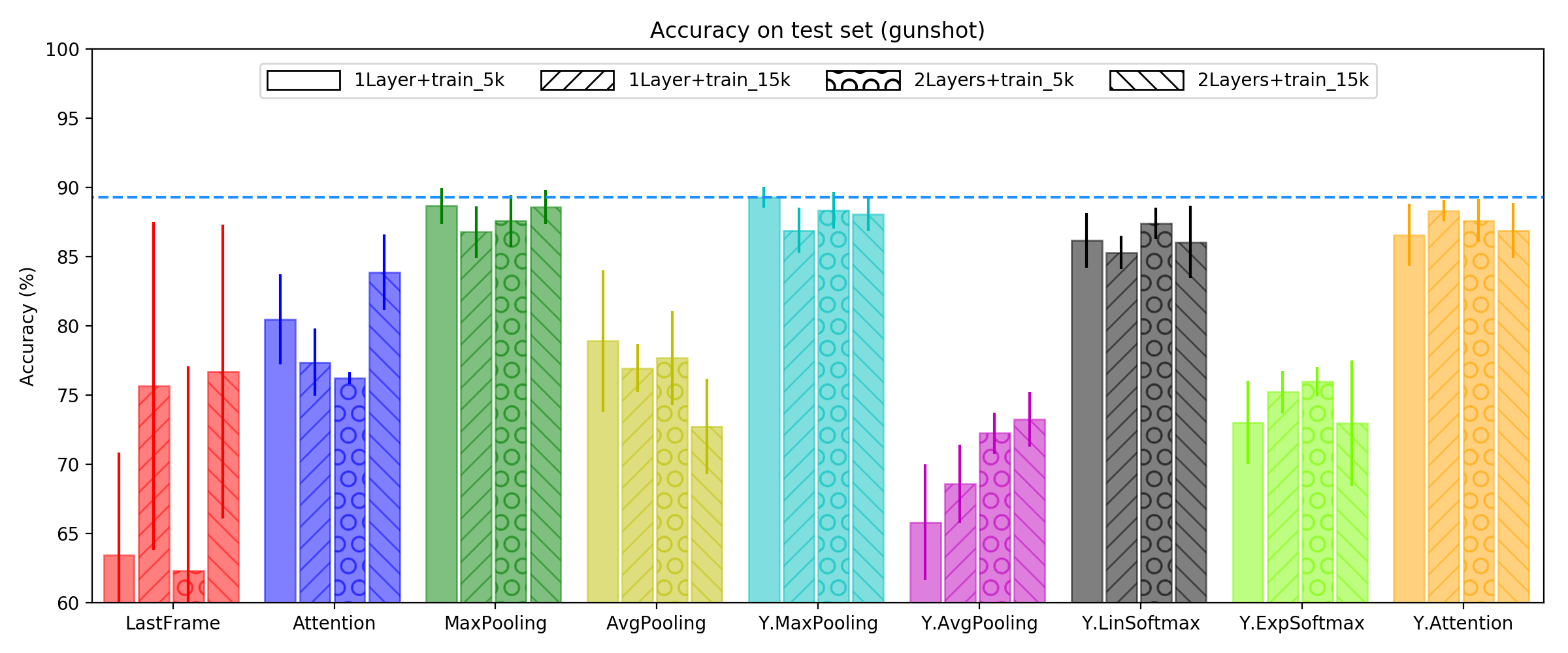}
        \caption{Gunshot}
        \label{fig:test_acc_gunshot}
    \end{subfigure}
    \vspace{-0.1in}
    \caption{Accuracies on the test set for LSTM models with different pooling methods. Each bar represents an average of five trials, and the error bar is the sample standard deviation of five trials. Note that names starting with ``\textit{Y.}'' are pooling methods on prediction, and the rest are pooling on feature. The horizontal dashed line is the best accuracy out of all setups.}
    \label{fig:acc_test}
    \vspace{-0.2in}
\end{figure}

\subsection{Pooling on prediction}
\label{s:poolingFeat}
Given the feature $\boldsymbol{x}_t$ generated by LSTM model at each frame $t$, we can generate the frame-level prediction $y_t$.
These frame-level predictions $\boldsymbol{Y} = [y_1,...,y_T]$ are fed into a pooling layer to generate the utterance-level prediction $y$.
We list the definition of five prediction-level pooling functions covered in~\cite{wang2019comparison} in Table~\ref{t:poolingPred}.
Max and average pooling generate the utterance-level prediction by simply taking the max and mean of frame-level predictions across all time frames.

The two softmax pooling functions are weighted sum of frame-level predictions, where the weights for linear softmax is the frame-level prediction itself ($y_t$), and the weights for exponential softmax is the exponential of frame-level prediction ($\exp{(y_t)}$).
These softmax pooling functions give the frames with high prediction values more influence on the utterance-level prediction.

The attention pooling function uses attention weights ($w_t$) learned from a dedicated dense layer within the network, as proposed in~\cite{wang2019comparison}.
Note that the attention weights here ($w_t$) are different from the one used in attention pooling on feature ($a_t$).
Unlike $w_t$, we don't have an extra dense layer to generate the attention weights for $a_t$, which is generated by sharing the weights with the dense layer generating utterance prediction $y$ as explained in~\cite{wang2018_IS}.


\begin{figure}[t!]
    \centering
    \hspace{-0.3in}
    \begin{subfigure}[b]{0.25\textwidth}
        \includegraphics[width=\textwidth]{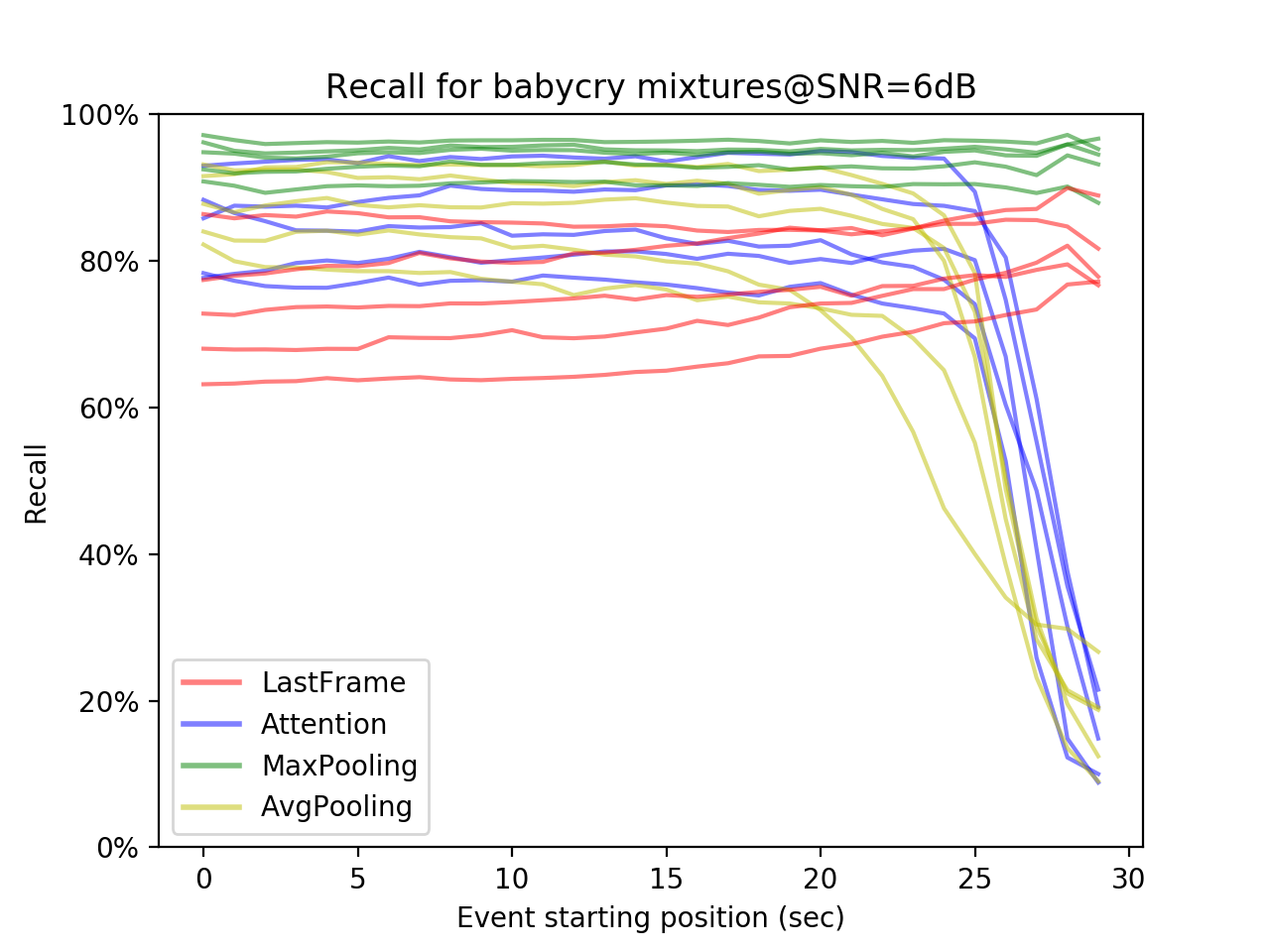}
        \label{fig:recall_BC_6dB_Fpool}
    \end{subfigure}
    \hspace{-0.25in}
    ~ 
    \begin{subfigure}[b]{0.25\textwidth}
        \includegraphics[width=\textwidth]{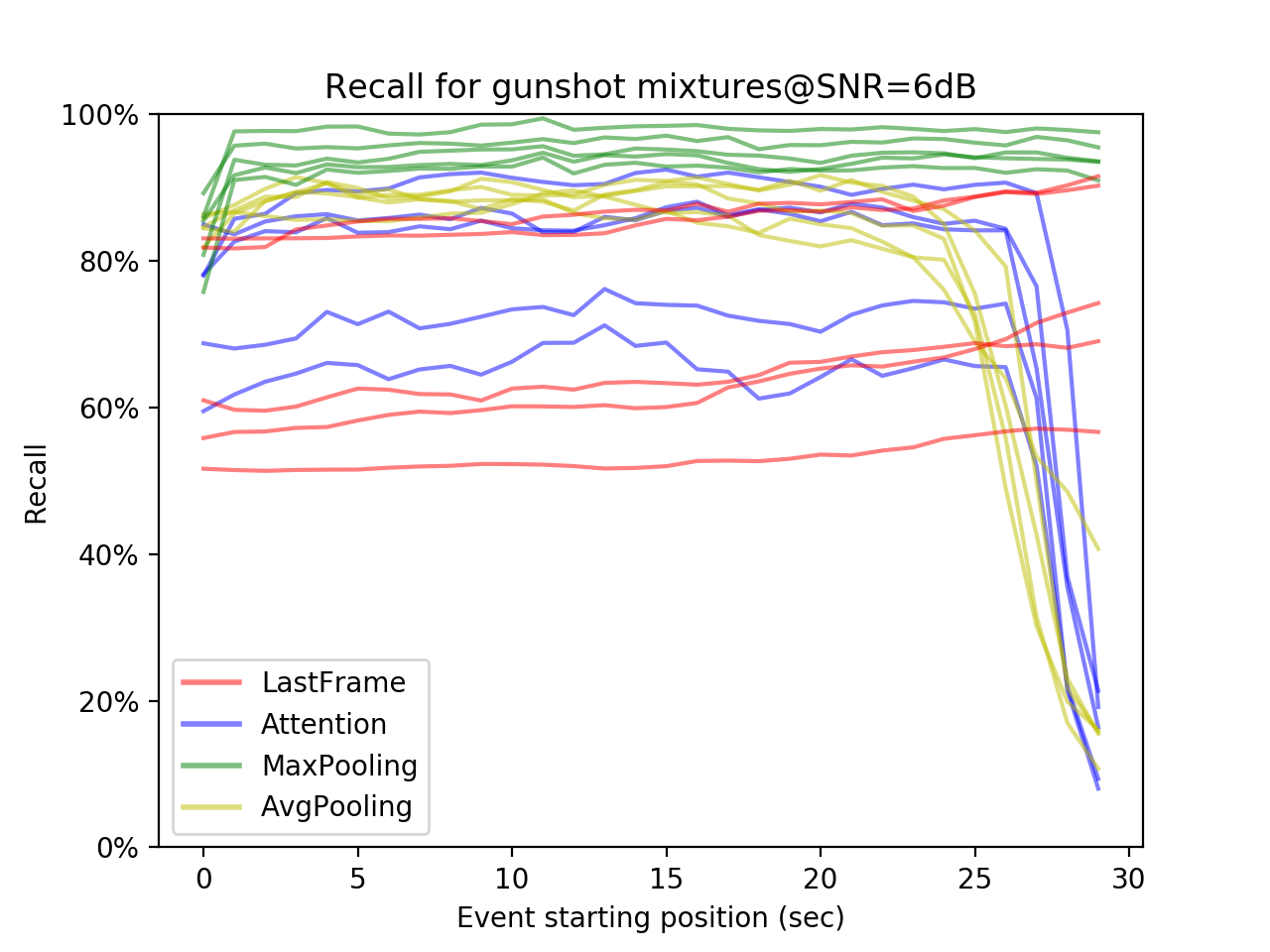}
        \label{fig:recall_GS_6dB_Fpool}
    \end{subfigure}
    \\
    \vspace{-0.15in}
    \hspace{-0.3in}
    \begin{subfigure}[b]{0.25\textwidth}
        \includegraphics[width=\textwidth]{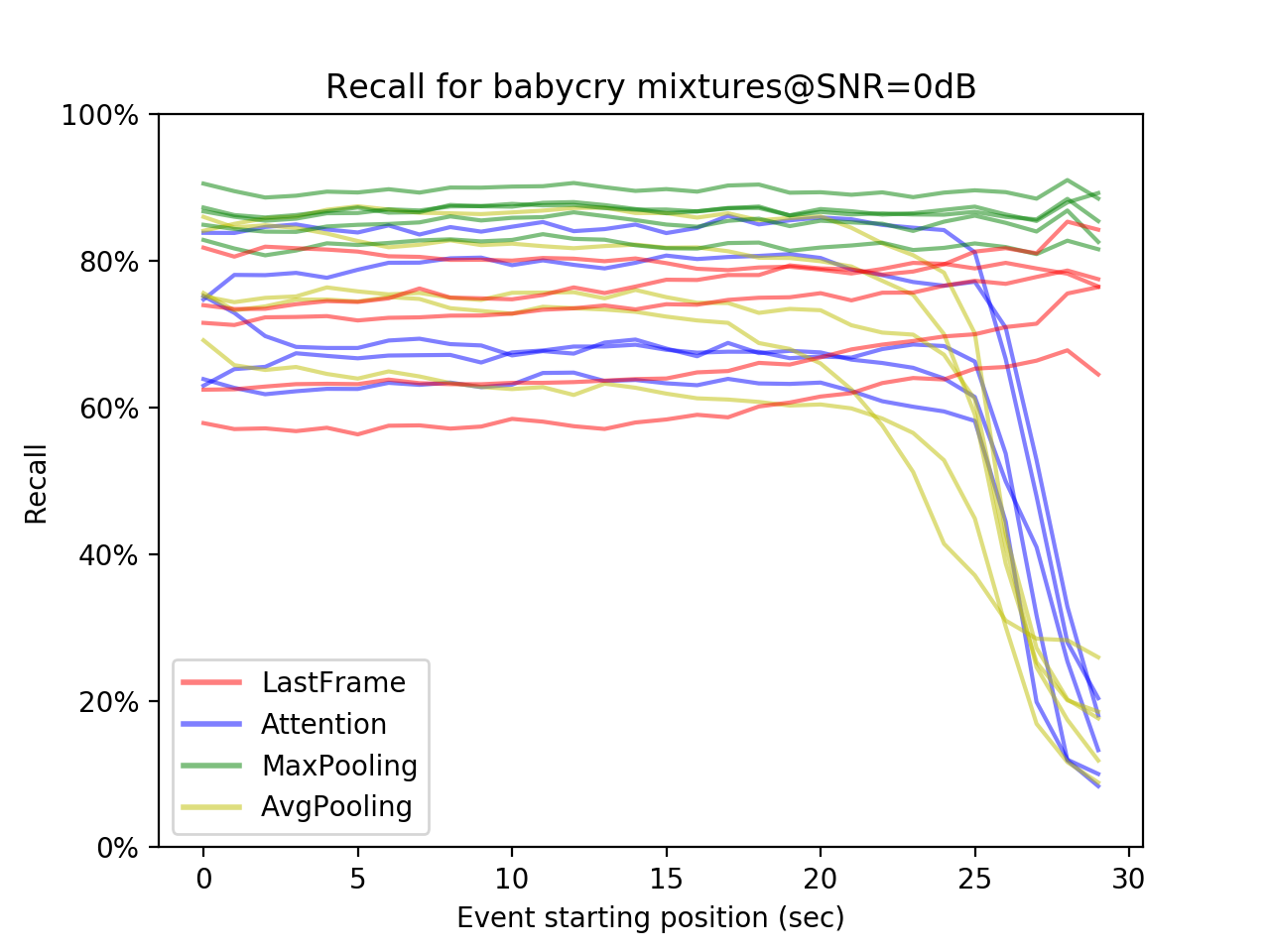}
        \label{fig:recall_BC_0dB_Fpool}
    \end{subfigure}
    \hspace{-0.25in}
    ~ 
    \begin{subfigure}[b]{0.25\textwidth}
        \includegraphics[width=\textwidth]{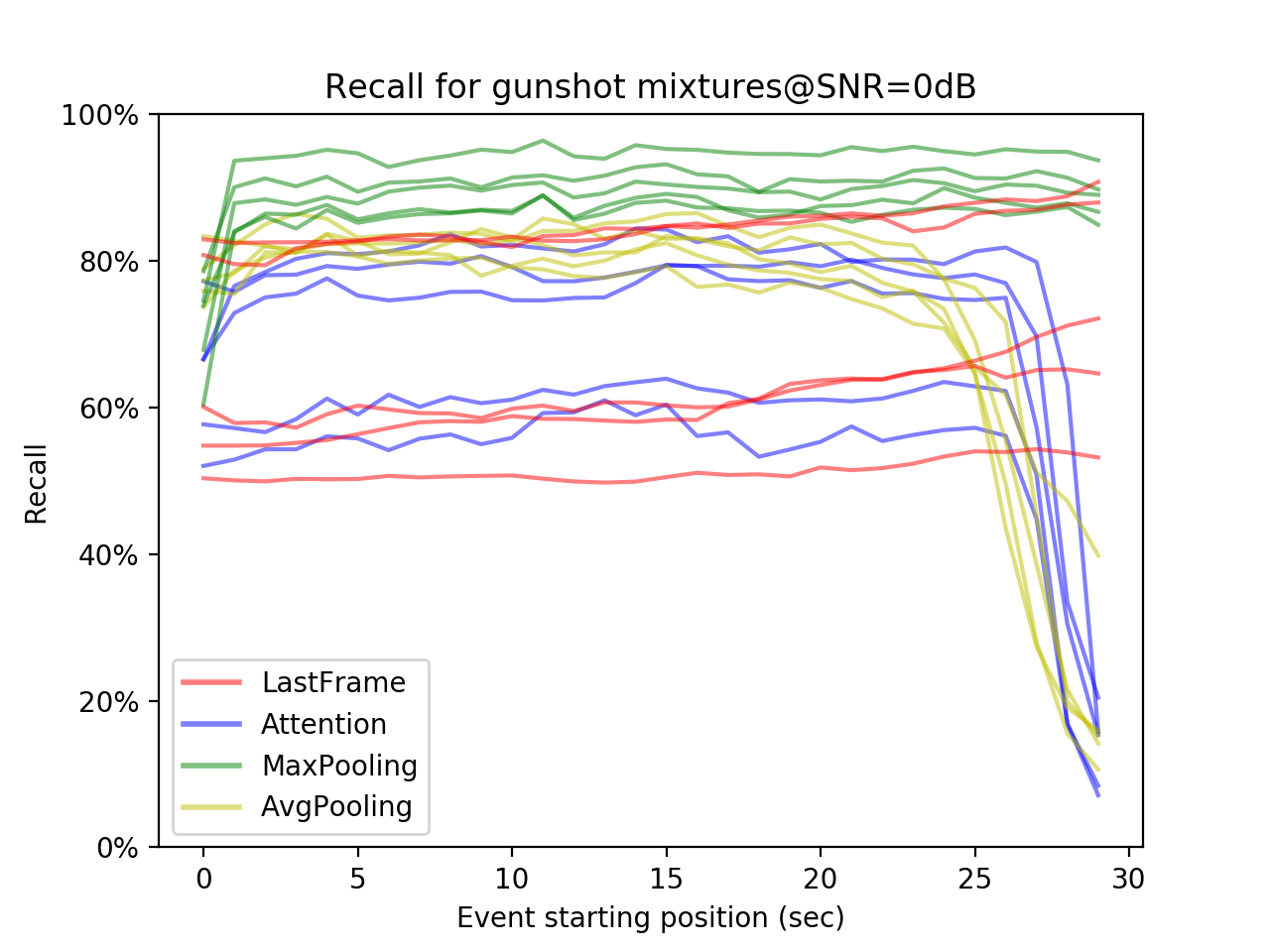}
        \label{fig:recall_GS_0dB_Fpool}
    \end{subfigure}
    \\
    \vspace{-0.2in}
    \caption{Recall rate for methods of pooling on feature for `babycry' abd `gunshot'. Mixtures used in this evaluation are generated by placing event segments at different event positions. Rows from top to bottom are mixtures with EBR of 6dB, 0dB respectively.}
    \label{fig:recall_Fpool}
    \vspace{-0.2in}
\end{figure}

\section{Experimental Setup}
\label{s:exp}


\subsection{Dataset}
\label{s:dataset}
We tested nine pooling methods on LSTM models using the dataset provided by DCASE 2017 Challenge task 2~\cite{DCASE2017challenge}: ``Detection of rare sound events.''
The task data consist of three isolated target events (baby crying, glass breaking, and gunshot) downloaded from freesound.org, and 30-second background sound clips including fifteen different audio scenes (bus, cafe, home, library, etc.) from TUT Acoustic Scenes 2016 dataset~\cite{TUT2016}.
Rare events are relatively short (averaged length shorter than 2 seconds in the test set) compared with the background clips.
A synthesizer provided by the challenge organizer is used to generate mixtures of target events and background sound clips at random onset time.
For each target event, we generate 5,000 or 15,000 training samples with event-to-background ratios (EBR) of -6, 0, 6dB.
These EBRs are chosen to match the data distribution of mixture clips provided in the dev/test sets.
For each generated training set, half of the utterances contain the target event, and the other half don't.
For the development and test sets, there are 500 mixtures provided by the challenge organizer in each set, and the ratio of containing event is 0.5 as well.
The evaluation metric we used for utterance-level binary classification is accuracy, and the threshold is set to 0.5 to all models in this work.

The synthesized mixtures are 30-second monaural audio with 44,100 Hz and 24 bits.
We use log filter bank energies (LFBEs) as the acoustic features in this work.
We decompose each mixture clip into a sequence of 25 ms frames with a 10 ms shift.
64 dimensional LFBEs are calculated for each frame, and we aggregate the LFBEs from all frames to generate the input spectrogram (3,000$\times${64}).

\subsection{Training Setup}
\label{s:training_setup}
We use models consist of one or two uni-directional LSTM layers, followed by different pooling methods to generate the utterance-level prediction. 
We set the number of units to 100 for all LSTM layers.
We use the loss of the development set as the criterion for model selection.
Adaptive momentum (ADAM)~\cite{ADAM} is used as the optimizer and the initial learning rate is set to 0.001. 
The size of mini-batch is set to 200 for the training set of 5k samples, and 600 for the training set of 15k samples. 
One exception is that for attention pooling on feature, we set it to 500 for model with one LSTM layer and 400 for model with two LSTM layers on the training set of 15k samples due to the memory constraints on the GPU.
We use Keras with Tensorflow backend on Tesla K80 GPUs to conduct the experiments.

\subsection{Dataset for Testing Sensitivity to Event Positions}
\label{s:dataset_sens}
In order to check the sensitivity to event positions for different pooling methods, we need a dataset containing mixtures with the target event placed at different timestamps.
We use the source data in the test set to generate such mixtures.
We first randomly selected 100 30-second background clips out of 387 clips from the test set.
For each event segment, we generate the mixtures by placing it at different positions $t \in [0, 1, ..., 29]$, and mix it with the background clip at three EBR of -6, 0, 6dB.
Depending on the length of event segment, it may generate up to 30 mixtures for each pair of event segment and background clip. 
In the test set, there are 61, 58, and 76 segments for `babycry', `glassbreak', and `gunshot' respectively.
We have generated roughly 1.76M 30-second mixtures ((61+58+76)$\times$100$\times$30$\times$3) for benchmarking the sensitivity of pooling methods to event positions.
We use the recall rate on mixtures as the evaluation metric since all mixtures are positive samples for the classification task.

\section{Experimental Results}
\label{s:results}

For each type of event, we first experiment nine pooling methods on two models (1 and 2 LSTM layers) and two sizes of the training set (5k and 15k). 
The accuracies on the test set are shown in Fig.~\ref{fig:acc_test}.
For architectures explored here with the best pooling (Y.MaxPooling), we observed that neither increasing the number of mixtures in the training set nor increasing the model complexity can further improve the accuracy.
This shows that models with a single LSTM layer have enough model complexity for DCASE 2017 task 2 dataset.
Therefore, we conduct our further analysis of different pooling methods on models with a single LSTM layer trained on the 5k dataset.

\begin{figure}[thb!]
    \centering
    \hspace{-0.3in}
    \begin{subfigure}[b]{0.25\textwidth}
        \includegraphics[width=\textwidth]{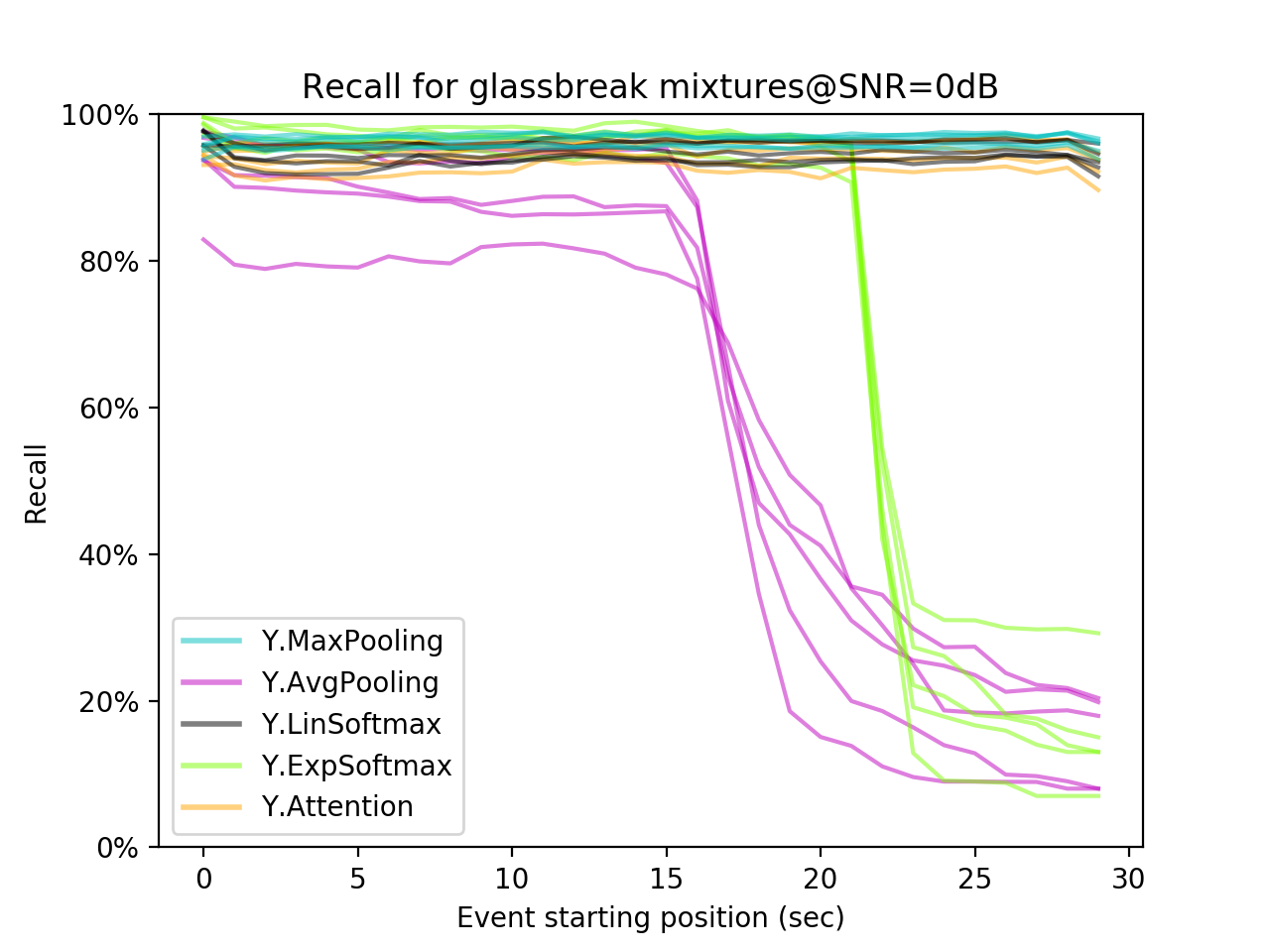}
        \label{fig:recall_GB_0dB_Ppool}
    \end{subfigure}
    \hspace{-0.25in}
    ~ 
    \begin{subfigure}[b]{0.25\textwidth}
        \includegraphics[width=\textwidth]{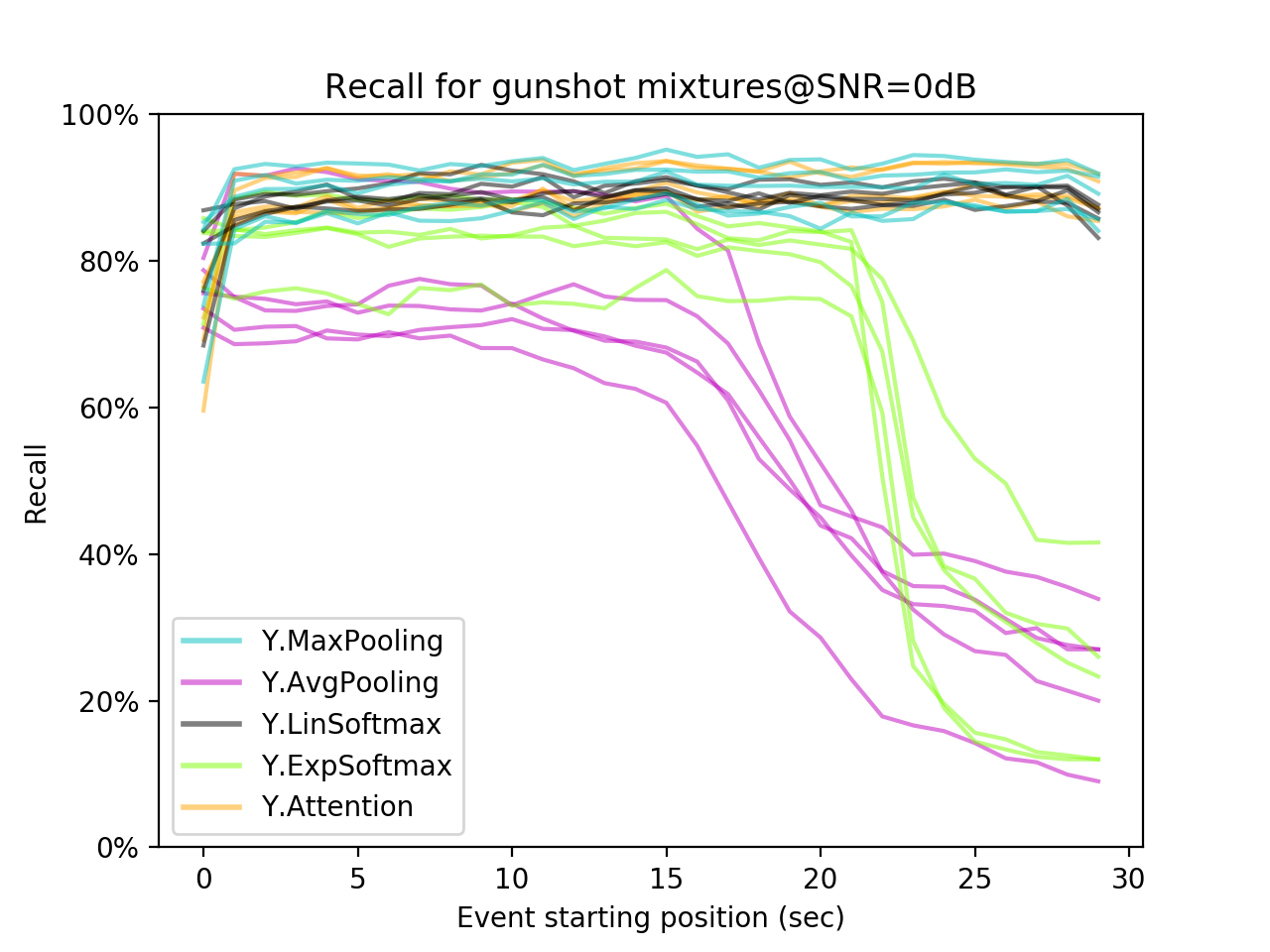}
        \label{fig:recall_GS_0dB_Ppool}
    \end{subfigure}
    \\
    \vspace{-0.15in}
    \hspace{-0.3in}
    \begin{subfigure}[b]{0.25\textwidth}
        \includegraphics[width=\textwidth]{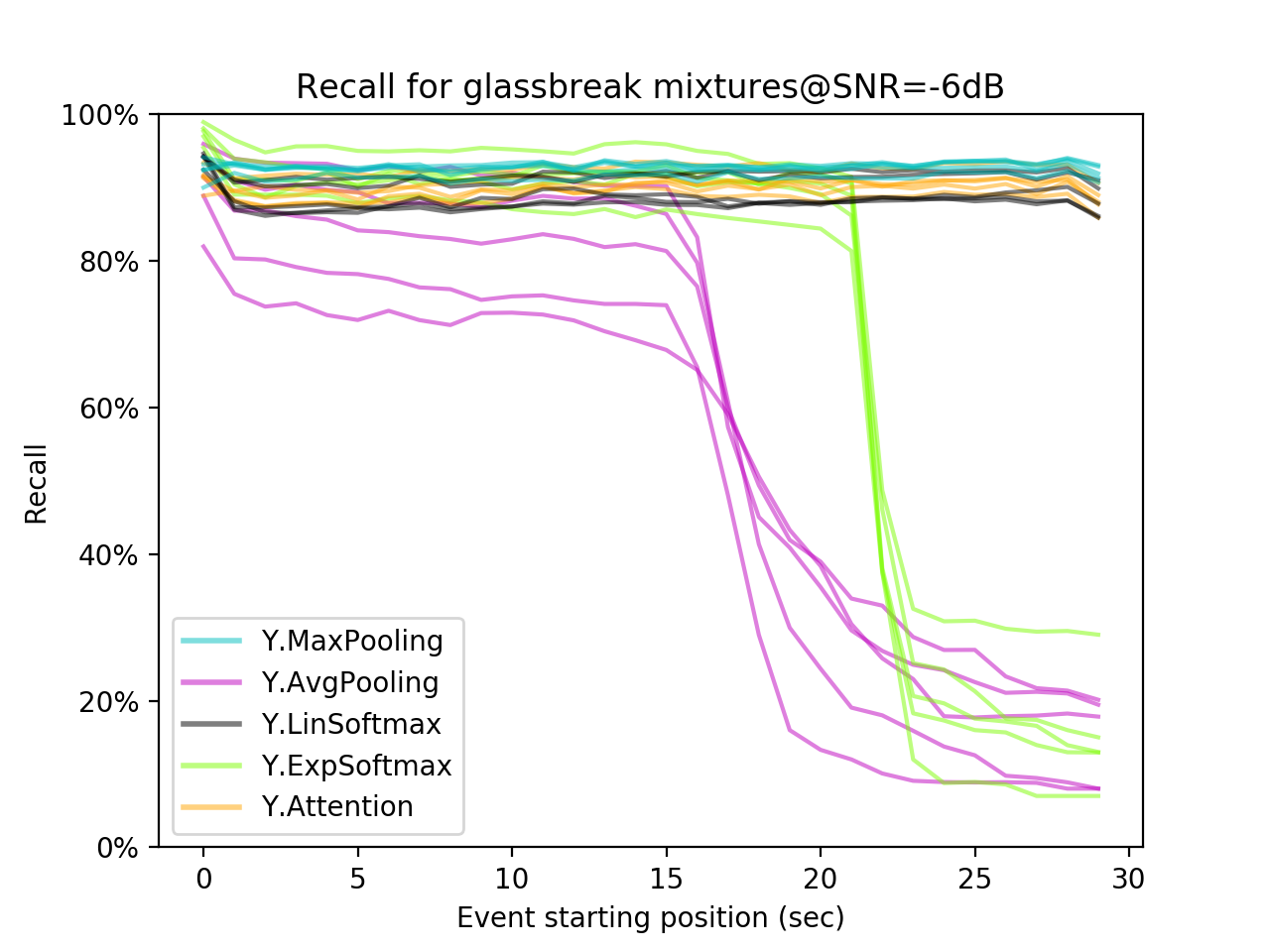}
        \label{fig:recall_GB_-6dB_Ppool}
    \end{subfigure}
    \hspace{-0.25in}
    ~ 
    \begin{subfigure}[b]{0.25\textwidth}
        \includegraphics[width=\textwidth]{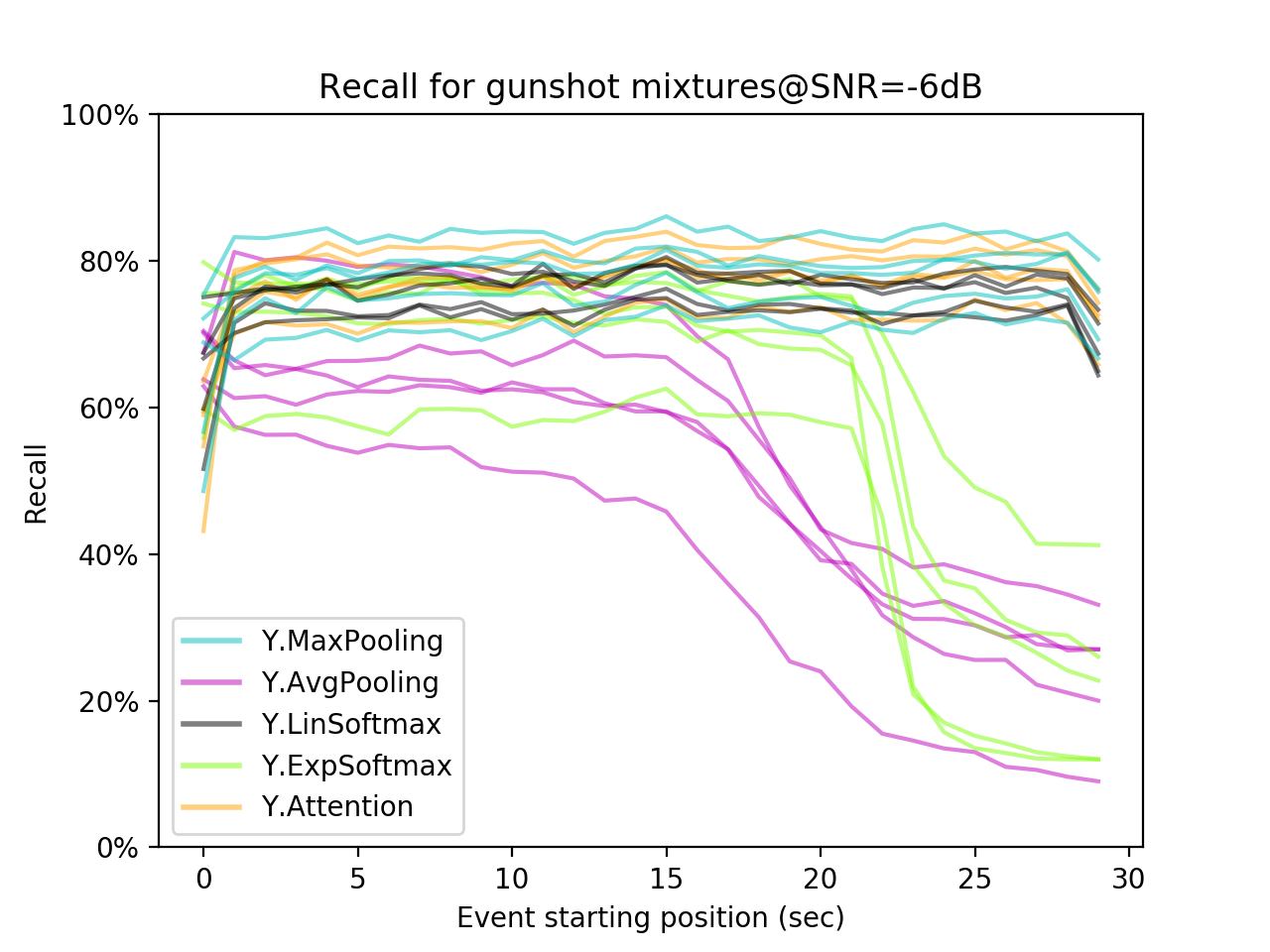}
        \label{fig:recall_GS_-6dB_Ppool}
    \end{subfigure}
    \\
    \vspace{-0.15in}
    \caption{Recall rate for methods of pooling on prediction for `babycry' and `gunshot'. Mixtures used in this evaluation are generated by placing event segments at different event positions. Rows from top to down are mixtures with EBR of 0dB, -6dB respectively.}
    \label{fig:recall_Ppool}
    \vspace{-0.15in}
\end{figure}

As shown in Fig.~\ref{fig:acc_test}, max pooling on prediction outperforms other eight pooling methods across three event types.
If we take the average of accuracies (1Layer+train\_5k) shown in Fig.~\ref{fig:acc_test} over three types of events, we can find the top-4 pooling methods ranked as following: 1) Y.MaxPooling (91.03\%), 2) MaxPooling (90.11\%), 3) Y.Attention (89.36\%), 4) Y.LinSoftmax (89.07\%).
This result is not consistent with the finding in~\cite{wang2019comparison}, where max pooling on prediction performs the worst for audio tagging.
We hypothesize that this inconsistency is due to the different characteristics between two datasets, and more details are covered in Sec.~\ref{s:dynamics}.




\subsection{Dynamics of LSTM Models on AEC}
\label{s:dynamics}

If the LSTM model is able to retain long-term memory over thousands of steps, it should be able to detect the event no matter where the event occurs within the 30-second utterance.
We investigate the dynamics of LSTM memory on AEC tasks by testing the models on mixtures with event segments placed at different timestamps.
For each model, we evaluate the recall rate on mixtures with an event segment placed at time $t$, and observe how the recall rate changes with respect to $t$.
Curves for pooling on feature methods (Fig.~\ref{fig:recall_Fpool}) and pooling on prediction methods (Fig.~\ref{fig:recall_Ppool}) were calculated on 1.7M mixtures synthesized with the setup described in Sec.\ref{s:dataset_sens}.

\noindent
\textbf{Memory retaining}
Li et al.~\cite{Li_2018_CVPR} showed that LSTM can only keep a mid-range memory (about 500-1,000 time steps).
To check if LSTM models have a similar memory forgetting issue on AEC, we can look at the red curves of `LastFrame' in Fig.~\ref{fig:recall_Fpool}.
For `babycry' and `gunshot' events, we can see the recall rate goes up with the increase of $t$, which shows that the LSTM model has higher chances to forget events happened close to the beginning of utterance.
This trend is universal across different EBR settings.
These results suggest that it is essential to choose a pooling method for LSTM models suitable for rare AEC.

\noindent
\textbf{Sensitivity to event positions}
From Fig.~\ref{fig:recall_Fpool} and Fig.~\ref{fig:recall_Ppool}, we can tell that the top 4 performing methods (Y.MaxPooling, MaxPooling, Y.Attention, and Y.LinSoftmax) are not sensitive to event positions.
Overall, the recall rates do not change a lot in respect of $t$.
Interestingly, there is a huge drop in recall rate when `gunshot' events are placed at $t=0$ for the top-4 methods.
We suspect that it is because of an issue of annotation for the onset time.
Models learned to identify `gunshot' sounds by detecting the sharp change in the amplitude of sounds.
The onset time of some `gunshot' events are labeled after the actual onset time, which means that there is no huge amplitude change in the event segment.
When these segments get placed at $t=0$, models are not able to detect them correctly.

As the magenta curves shown in Fig.~\ref{fig:recall_Ppool}, the top performing pooling method for audio tagging (Y.AvgPooling) reported in~\cite{wang2019comparison} is very sensitive to event position. 
If the event is placed at after half of the mixture ($t>=15$), the recall rate decreases quickly with respect to $t$.
This verifies our hypothesis that different characteristics of datasets lead to findings inconsistent with~\cite{wang2019comparison}.
In ~\cite{wang2019comparison}, the event length is much longer than the rare events in our setup and it covers a huge part of an utterance, which means the onset time of the event is close to the beginning of the utterance.
Y.AvgPooling works for that case since sensitivity to event position is not an issue anymore, and it can also avoid some noises brought by using max pooling.
On the contrary, Y.AvgPooling is not suitable for rare AEC due to that the event length is short than 2 seconds, and the sensitivity to event position matters a lot.


\begin{figure}[t!]
    \centering
    \hspace{-0.3in}
    \begin{subfigure}[b]{0.272\textwidth}
        \includegraphics[width=\textwidth]{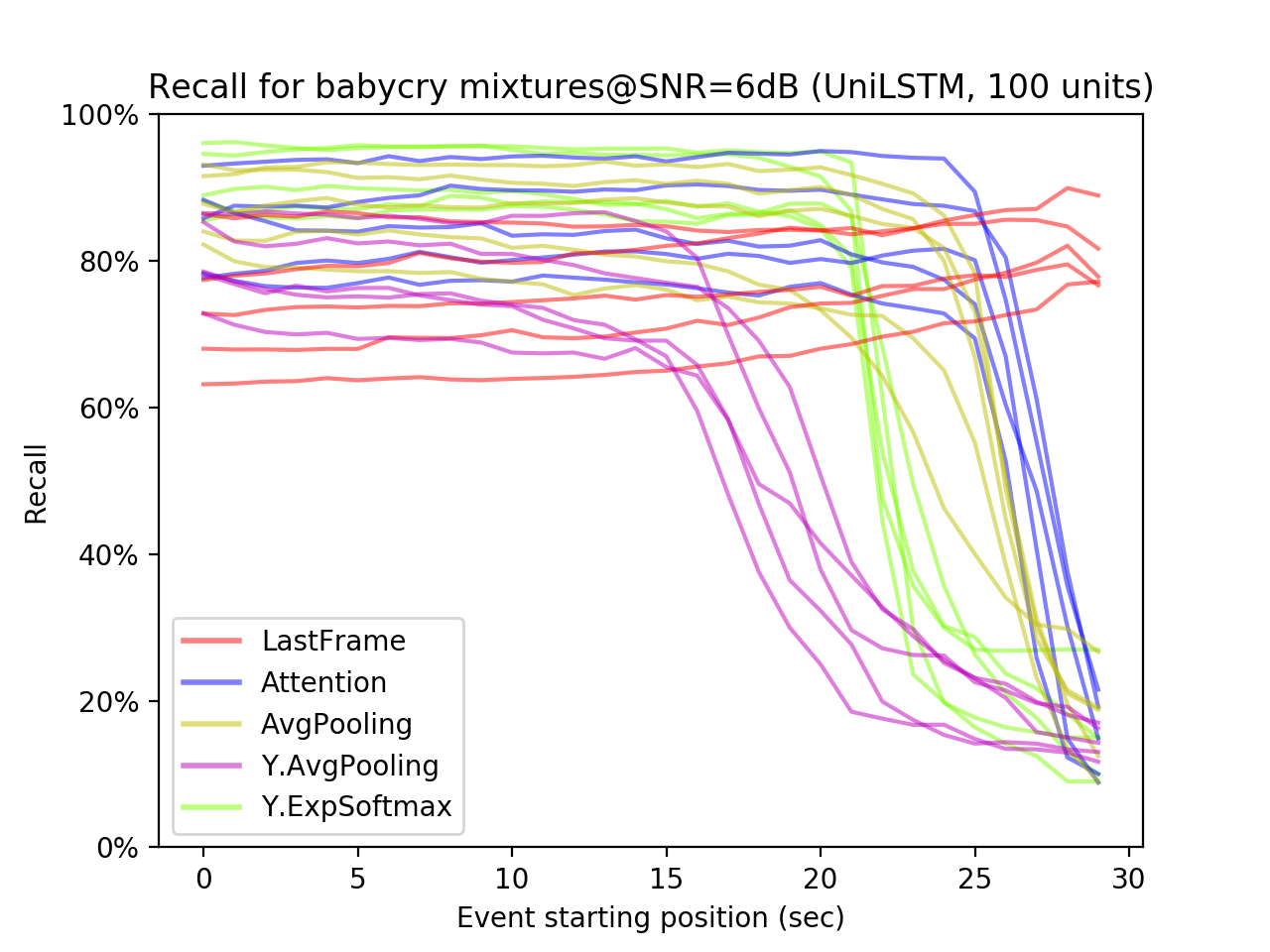}
        \caption{Uni(100 units)}
        \label{fig:recall_UniLSTM100}
    \end{subfigure}
    \hspace{-0.29in}
    ~
    \begin{subfigure}[b]{0.272\textwidth}
        \includegraphics[width=\textwidth]{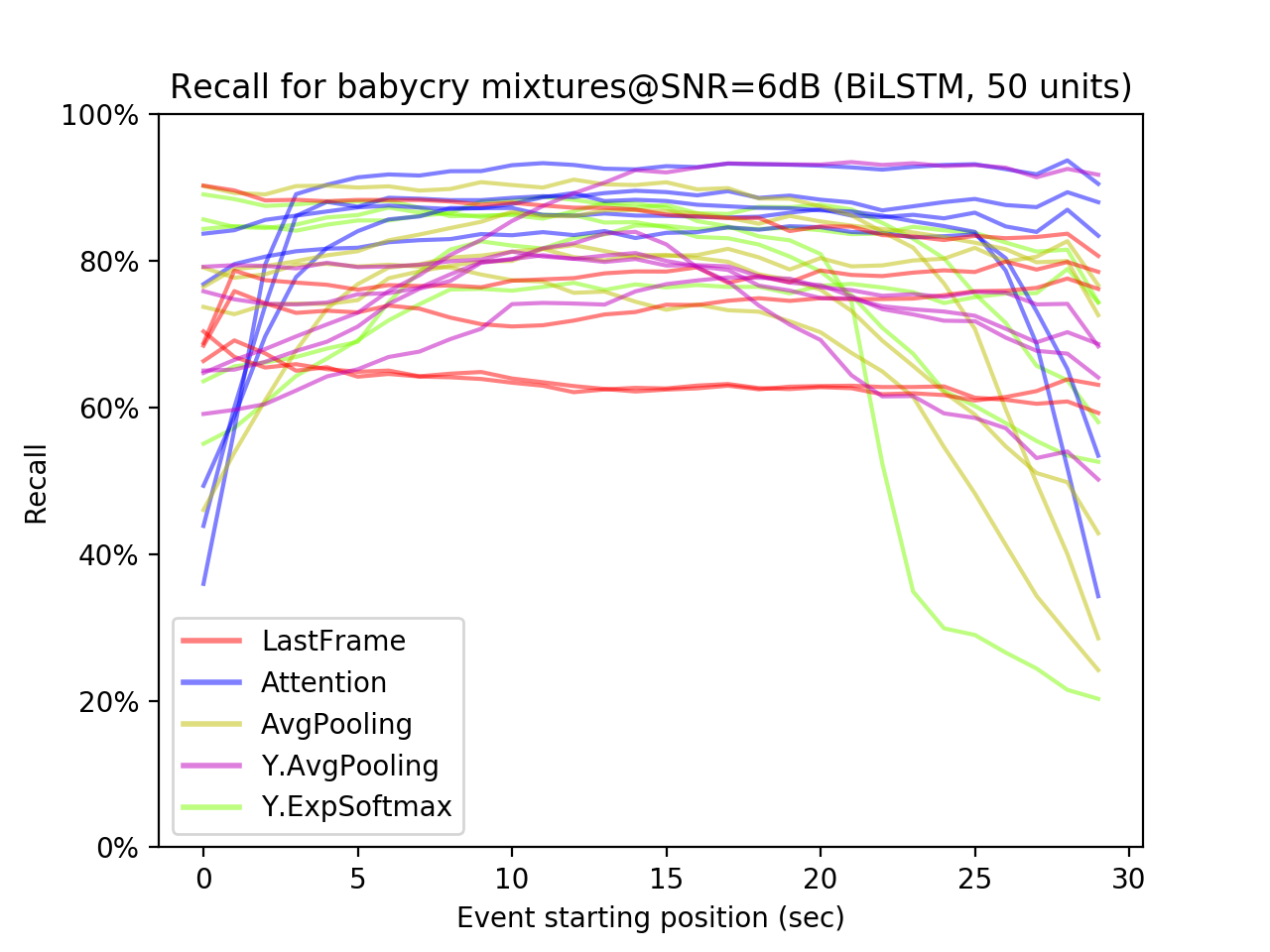}
        \caption{Bi(50 units)}
        \label{fig:recall_BiLSTM_50}
    \end{subfigure}
    \vspace{-0.12in}
    \caption{We applied 5 pooling methods sensitive to event positions on uni-directional (Uni) and bi-directional (Bi) LSTM models. Recall rates are calculated on mixture clips of `babycry' with 6dB SNR.}
    \label{fig:BiLSTM}
    \vspace{-0.25in}
\end{figure}

\noindent
\textbf{Mitigation of sensitivity}
After observing the sensitivity to event positions for certain pooling methods, we are looking for a solution to mitigate this effect.
A straight forward idea is to apply these pooling methods to a bi-directional LSTM model.
We have trained a set of bi-directional LSTM models with 50 units in each direction. This setup generates feature maps with 100 units at each frame, which is the same as the default model (100 units uni-directional LSTM model). 
For each pooling method, we trained 5 models to reduce the randomness during the training.
We tested these models on mixture clips of `babycry' event with 6dB SNR and the recall rates are shown in Fig.~\ref{fig:BiLSTM}.
As shown in Fig.~\ref{fig:recall_BiLSTM_50}, the sensitivity to event positions is reduced significantly by using bi-directional LSTM.
Compared with Fig.~\ref{fig:recall_UniLSTM100}, the worst recall rate of five trials has improved from 10\% to 50\% for Y.AvgPooling, from 10\% to 35\% for Attention, from 10\% to 25\% for AvgPooling. 

\section{Conclusion}
\label{s:conclusion}

We have benchmarked nine different pooling methods for LSTM AEC models on task 2 of the DCASE 2017 challenge, with 1.7M mixtures generated to evaluate utterance-level classification accuracy and sensitivity to event positions.
We found that max pooling on the prediction level (Y.MaxPooling) is the best performing method in terms of classification accuracy, and it is also robust to event positions.
This observation is different from the finding in~\cite{wang2019comparison}, where max pooling on prediction is the worst performing method in audio tagging task.
We hypothesize that this inconsistency is due to the different characteristics between two datasets as discussed in Sec.~\ref{s:dynamics}.
We also explored using bi-directional LSTM models to mitigate the sensitivity issue for certain pooling methods.
To authors' best knowledge, this is the first work focusing on LSTM memory dynamics for AEC tasks.

\bibliographystyle{IEEEbib}
\bibliography{refs}

\end{document}